# Intrinsic Raman Modes and Crystal Symmetry of Altermagnetic α-MnTe

Yunpeng Cui[1#], Shuai Zhang[2#], Jia-Liang Xie[3#], Jinxuan Liang[1#], Hao Lin[1], Miao Qin[1], Jiawei Shao[1], Gongling Chen[1], Yidian Li[1], Yixuan Luo[4], Deng Hu[5,6], Bingru Lu[7], Hao Li[7], Xiang Li[7], Tiantian Liu[7], Haodong Yu[1], Hongyun Zhang[1], Shuyun Zhou[1,8], Zhiwei Wang[5,6,9], Yanfeng Guo[4,10], Tiantian Zhang[2], Miao-Ling Lin[3], Ping-Heng Tan[3,*], Feng Jin[11,*], Luyi Yang[1,8,*]

*[1]State Key Laboratory of Low Dimensional Quantum Physics, Department of Physics, Tsinghua University, Beijing 100084, China.*

*[2]Institute of Theoretical Physics, Chinese Academy of Sciences, Beijing 100190, China.*

*[3]State Key Laboratory of Semiconductor Physics and Chip Technologies, Institute of Semiconductors, Chinese Academy of Sciences, No. A35, East Qinghua Road, Haidian District, Beijing 100083, China.*

*[4]State Key Laboratory of Quantum Functional Materials, School of Physical Science and Technology, ShanghaiTech University, Shanghai 201210, China.*

*[5]Centre for Quantum Physics, Key Laboratory of Advanced Optoelectronic Quantum Architecture and Measurement (MOE), School of Physics, Beijing Institute of Technology, Beijing 100081, China.*

*[6]Beijing Key Laboratory of Quantum Matter State Control and Ultra-Precision Measurement Technology, Beijing Institute of Technology, Beijing 100081, China.*

*[7]School of Physical Science and Technology, Key Laboratory of Quantum Theory and Applications of MoE, Lanzhou University, Lanzhou 730000, China.*

*[8]Frontier Science Center for Quantum Information, Beijing, People's Republic of China.*

*[9]Marine Science and Technology Domain, Beijing Institute of Technology, Zhuhai 519088, China.*

*[10]ShanghaiTech Laboratory for Topological Physics, ShanghaiTech University, Shanghai 201210, China.*

*[11]Beijing National Laboratory for Condensed Matter Physics, Institute of Physics, Chinese Academy of Sciences, Beijing 100190, China.*

*[#] These authors contributed equally to this work.*

*[*] Email: phtan@semi.ac.cn, jinfeng@iphy.ac.cn, luyi-yang@mail.tsinghua.edu.cn*

## Abstract

**Crystal symmetry governs the physical properties of materials and provides the framework for interpreting their microscopic excitations and macroscopic responses. In altermagnets, the crystal symmetries that relate opposite-spin sublattices give rise to momentum-dependent spin splitting despite compensated magnetization. Determining the real crystal symmetry is therefore essential for understanding the intrinsic electronic and magnetic properties. For the altermagnetic semiconductor α-MnTe, conflicting Raman assignments have left both its intrinsic lattice excitations and its crystal symmetry unresolved. Here we combine high-resolution Raman and second-harmonic generation (SHG) spectroscopy with first-principles calculations to distinguish intrinsic lattice excitations from Te-derived spectral contributions and to examine bulk inversion symmetry of α-MnTe. We show that several prominent Raman features previously assigned to α-MnTe instead originate from segregated Te clusters or precipitates. The remaining intrinsic Raman response consists of a single Raman-active mode assigned to the calculated $E_{2g}$ phonon, consistent with the centrosymmetric $D_{6h}$ symmetry. The SHG response agrees with surface electric-dipole and bulk electric-quadrupole contributions expected for a centrosymmetric $D_{6h}$ crystal, supporting the preservation of bulk inversion symmetry in α-MnTe. Our findings revise the Raman assignments in α-MnTe and provide a basis for interpreting its altermagnetic properties within the appropriate crystal-symmetry framework.**

Altermagnets form a class of magnetic materials distinct from both ferromagnets and conventional antiferromagnets. Like collinear antiferromagnets, they possess compensated order with vanishing net magnetization, yet their electronic bands can be spin split and spin polarized—features typically associated with ferromagnets [1-9]. Electrically accessible magnetoresistive signals and negligible stray fields make altermagnets promising candidates for high-density, field-robust spintronic memory devices [10, 11].

α-MnTe (hereafter referred to as MnTe) is a semiconducting altermagnet with a Néel temperature of $T_N \approx$ 307 K. It has emerged as an important platform for altermagnetism research, displaying a spontaneous anomalous Hall and magneto-optical Kerr response [12-14], nonrelativistic spin-split bands [15], nanoscale altermagnetic domains [16], chiral-split magnons [17], and ultrafast spin and charge dynamics [18-21]. Despite this progress, the fundamental crystal symmetry underlying these phenomena remains unresolved, particularly whether inversion symmetry is preserved.

Crystal structure determines the number and symmetry of first-order Raman-active modes, making Raman spectroscopy a sensitive probe of crystal symmetry. For the widely assumed centrosymmetric $D_{6h}$ ($6/mmm$) structure, group theory allows only a single first-order Raman-active phonon mode with $E_{2g}$ symmetry, with a calculated frequency of about 98 cm$^{-1}$. However, experiments have reported additional features near 122, 141, 175, and 265 cm$^{-1}$, as summarized in Table 1 [22-29]. These features have been assigned to first- or higher-order phonon scattering [22-25], two-magnon scattering [22, 23], an inversion-breaking $D_{3h}$ ($\bar{6}m2$) distortion [26], an $MnTe_2$ secondary phase [24, 29], or a plasmonic excitation [27]. Separating intrinsic MnTe modes from extrinsic contributions is therefore essential for reliable Raman assignments and clarifying the crystal symmetry that governs its physical properties.

In this work, we combine high-resolution Raman and second-harmonic-generation (SHG) spectroscopy with first-principles calculations to identify the intrinsic Raman-active phonons in MnTe and determine its crystal symmetry. Comparisons between low-temperature Raman spectra of MnTe and a Te single crystal, together with helicity- and spatially resolved Raman measurements, distinguish intrinsic MnTe modes from Te-derived features. First-principles calculations support assigning a single intrinsic Raman-active $E_{2g}$ mode to centrosymmetric $D_{6h}$ MnTe. Both our experiments and calculations reveal a substantially stronger Raman response from Te than from MnTe. Complementary SHG measurements are also consistent with $D_{6h}$ symmetry, with contributions from surface electric-dipole and bulk electric-quadrupole processes.

**Table 1. Comparison of reported Raman assignments with the findings of this work.** The listed Raman shifts are approximate room-temperature values.

| Frequency (cm$^{-1}$) | Previous reports | This work |
|---|---|---|
| 93 | Intrinsic MnTe $E_{2g}$ phonon [24];<br>Te-related phonon [28]. | Splits into three peaks at 1.8 K:<br>P1: Te $E^1$ phonon;<br>P2: polarization dependence and frequency consistent with the MnTe $E_{2g}$ mode;<br>P3: maybe related to strain. |

| Frequency ($cm^{-1}$) | Previous reports | This work |
|---|---|---|
| 122 | Intrinsic MnTe excitation beyond Γ-point optical phonons [24];<br>Distortion-activated $E'$ (TO) phonon [26];<br>Te-related phonon [27, 28]. | Te $A_1$ phonon. |
| 141 | Intrinsic MnTe excitation beyond Γ-point optical phonons [24];<br>Distortion-activated $E'$ (LO) phonon [26];<br>Te-related phonon [27, 28]. | Te $E^2$ phonon. |
| 175 | MnTe $E_{2g}$ phonon [22, 23];<br>Distortion-activated $A_1{}'$ phonon [26];<br>$MnTe_2$ impurity phonon [24, 29];<br>Plasmonic excitation [27];<br>Laser-damaged MnTe [29]. | Not observed. |
| 265 | Two-magnon scattering [23];<br>Two-phonon continuum [26]. | Te-related scattering. |

Figure 1(a) illustrates the hexagonal NiAs-type structure of MnTe. Along the $c$ axis, Mn and Te planes stack alternately; within the *ab* plane, both Mn and Te atoms form hexagonal lattices. Below $T_N$, the magnetic moments associated with Mn atoms align ferromagnetically within each Mn layer, while adjacent Mn layers couple antiferromagnetically along $c$, forming a compensated collinear spin texture [33-35].

Figure 1(b) shows the phonon dispersion curves of MnTe calculated using density functional theory, while Fig. 1(c) summarizes the frequencies and symmetry assignments of the Γ-point optical phonons. Details of the computational methods are provided in the Methods section. Within the centrosymmetric $D_{6h}$ structure, the doubly degenerate $E_{2g}$ mode is Raman-active, with a calculated frequency of 98.69 cm$^{-1}$. The $E_{1u}$ and $A_{2u}$ modes are infrared-active, with calculated frequencies of 126.54 and 129.11 cm$^{-1}$, respectively, whereas the remaining optical modes are silent.

Raman measurements were performed in a backscattering geometry on the (0001) surface of MnTe single crystals using 632.8 nm excitation. The sample was mounted in a cryostat with a base temperature of 1.8 K. The spectral resolution of approximately 0.35 cm$^{-1}$ was sufficient to distinguish closely spaced low-energy features. Unless otherwise specified, all Raman data were acquired with linearly polarized incident light and without polarization analysis of the scattered light. Further details of the setup can be found in the Methods section.

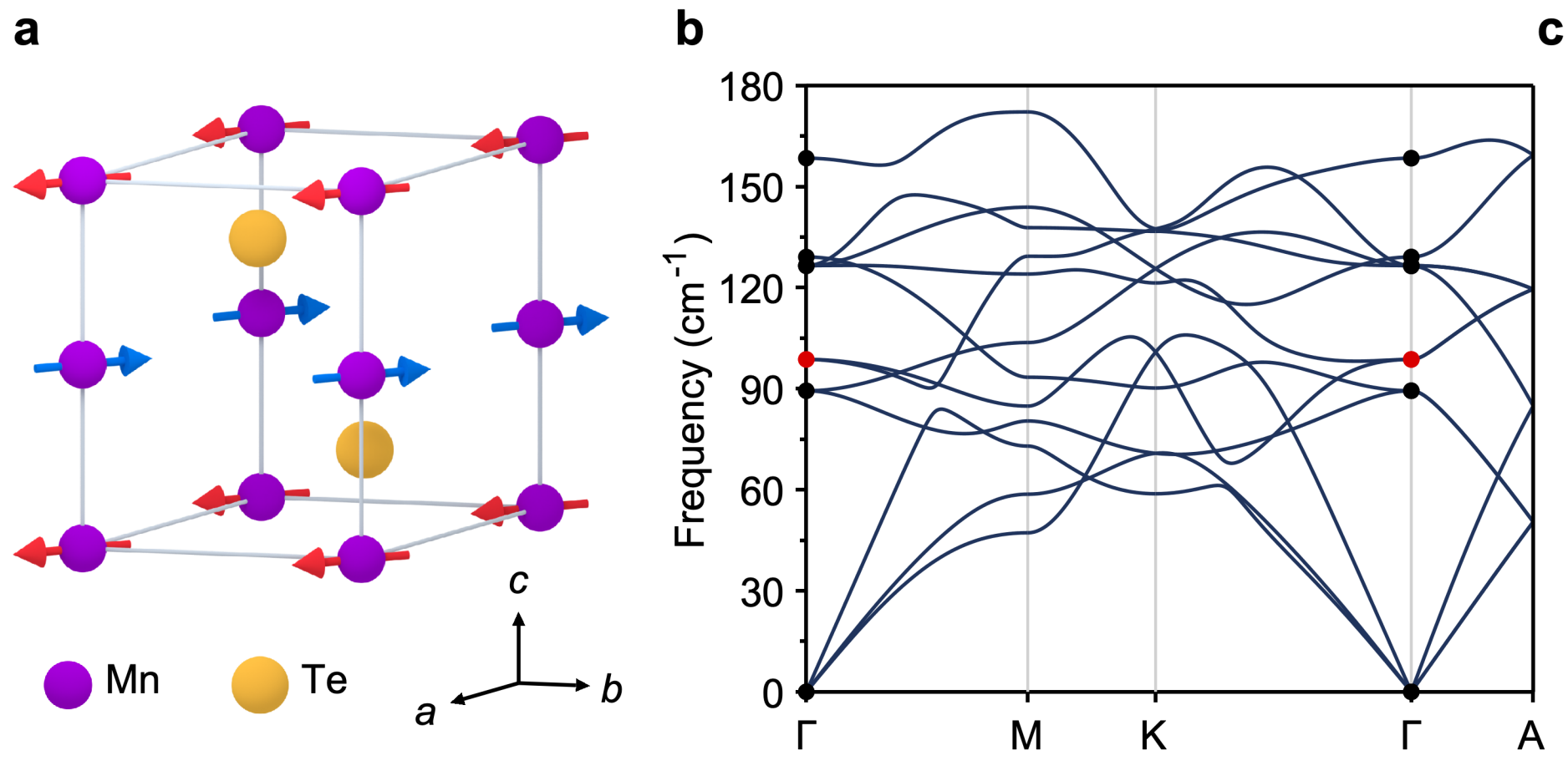

c

| Frequency (cm$^{-1}$) | Mode | Optical activity |
|---|---|---|
| 89.39 | $E_{2u}$ | Silent |
| 98.69 | $E_{2g}$ | Raman-active |
| 126.41 | $B_{2g}$ | Silent |
| 126.54 | $E_{1u}$ | IR-active |
| 129.11 | $A_{2u}$ | IR-active |
| 158.43 | $B_{1u}$ | Silent |

**Fig. 1 Structures and calculated phonon dispersion curves of MnTe.** (a) Crystal and magnetic structures of MnTe. (b) Calculated phonon dispersion relations along the Γ–M–K–Γ–A path. Dots mark the phonon frequencies at Γ, with the Raman-active $E_{2g}$ mode highlighted in red. (c) Calculated frequencies, symmetry labels, and optical activities of the Γ-point optical phonon modes.

Room-temperature Raman spectra of MnTe exhibit peaks near 93, 122, and 141 cm$^{-1}$ [Fig. 2(a)] and 265 cm$^{-1}$ [Fig. S4]. Although a 175 cm$^{-1}$ peak reported in earlier work has been ascribed either to an intrinsic MnTe mode [22, 23] or to an $MnTe_2$ secondary phase [24, 29], we find no detectable peak around this frequency. On cooling to 1.8 K, the Raman peaks harden and narrow, revealing five sharp peaks labeled P1–P5 in the range from 80 to 160 cm$^{-1}$ [Fig. 2(a)]. In particular, the room-temperature feature near 93 cm$^{-1}$ resolves into P1–P3. Helicity-resolved measurements further reveal a splitting in the P1 peak frequency between the RL and LR configurations [Fig. 2(b)], where the first and second letters denote the polarizations of the incident and scattered light, respectively, with R (L) denoting right- (left-) circular polarization. Lorentzian fits give a center-frequency difference of 1.07 cm$^{-1}$, well above our spectral resolution of approximately 0.35 cm$^{-1}$.

Such a splitting is not expected for the $E_{2g}$ ($E'$) phonons in the $D_{6h}$ ($D_{3h}$) model: three vertical mirrors enforce degeneracy between branches of opposite pseudoangular momentum [36]. Te crystal, by contrast, belongs to the chiral point group $D_3$ (32), which lacks vertical mirrors and permits an RL–LR splitting at finite wavevector [30, 36]. The agreement of P1 with the Te $E^1$ chiral phonon mode in frequency, helicity dependence, and splitting magnitude supports a Te-derived origin [30-32]. Furthermore, the P4 and P5 peaks at 122 and 141 cm$^{-1}$ are also in line with the $A_1$ and $E^2$ modes in Te, respectively [30-32]. By contrast, the P2 peak agrees with the calculated MnTe $E_{2g}$ mode in both frequency [Fig. 1(c)] and helicity dependent Raman selectivity [Fig. S6(a)].

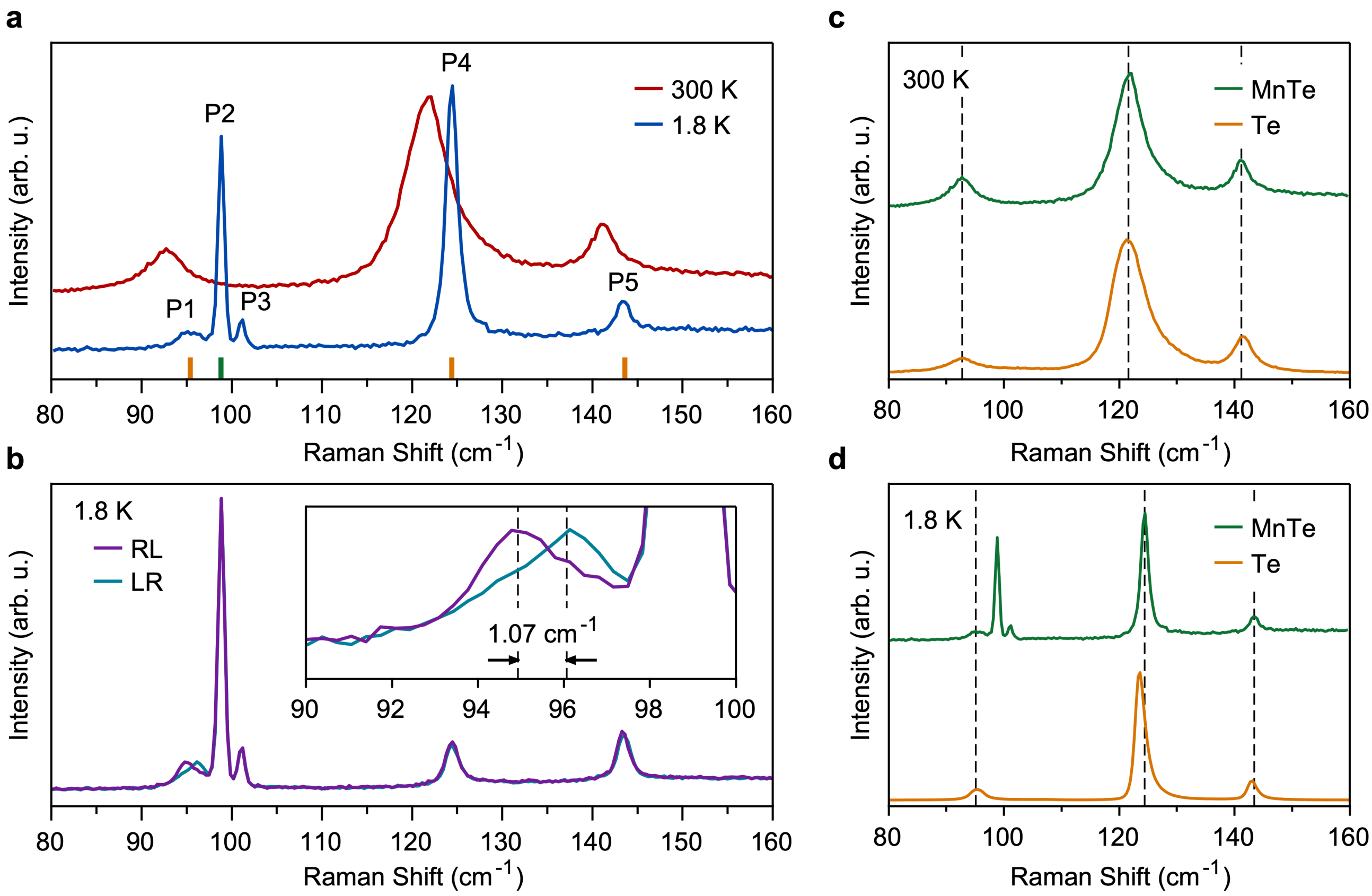

**Fig. 2 Raman spectra of MnTe and Te.** (a) Raman spectra of MnTe measured at 1.8 and 300 K, vertically offset for clarity. The five low-temperature peaks are labeled P1–P5; green and orange markers indicate peaks assigned to MnTe and Te, respectively. (b) Raman spectra of MnTe in the RL and LR configurations at 1.8 K. The inset highlights P1; the dashed lines mark the Lorentzian-fitted center frequencies, separated by 1.07 cm$^{-1}$. (c,d) Raman spectra of MnTe and Te at 300 K and 1.8 K, respectively. The spectra are normalized and vertically offset for clarity. Dashed vertical lines serve as guides for comparing the Te-related features.

To directly test the Te-derived assignments, we compared Raman spectra of MnTe(0001) and a Te(0001) single crystal, acquired at 300 and 1.8 K under identical experimental conditions. At 300 K, both samples show three peaks with the same frequencies [Fig. 2(c)]. At 1.8 K, we observe three peaks in Te at the same frequencies as P1, P4 and P5 [Fig. 2(d)]. In addition, the Te control sample also reproduces the 265 cm$^{-1}$ peak [Fig. S4], which was previously attributed to two-magnon scattering in MnTe [23]. This observation further identifies this peak as Te-derived rather than an intrinsic MnTe two-magnon response.

Furthermore, we also compared polarization-resolved Raman spectra of MnTe(0001) and Te(0001), acquired at 1.8 K under identical experimental conditions [Figs. S5(a)–S5(b)]. P1–P3 and P5 are most prominent in the cross-circular spectra, but are strongly suppressed in the co-circular spectra, while P4 follows the opposite trend. The Te control sample reproduces not only the frequencies of P1, P4, and P5 but also their polarization selection rules and relative intensities. The small frequency shifts relative to the Te crystal may reflect residual strain imposed by the MnTe host.

In addition, spatially resolved Raman spectra at 1.8 K show similar spatial intensity variations of P1, P4, and P5 after normalization to the P2 peak intensity [Fig. 3(a)]. Their linewidths also vary with position. Notably, these three peaks are weak or unresolved at Point 1, suggesting the presence of a relatively pristine local MnTe domain.

In short, our temperature-dependent, polarization-resolved, and spatially resolved Raman measurements of MnTe(0001) and Te(0001) support a common Te-related origin for P1, P4, and P5, which correspond to the $E^1$, $A_1$, and $E^2$ phonons of Te, respectively. At 1.8 K, the $A_1$ peak in the Te single crystal is approximately 66 times higher than the intrinsic P2 peak in MnTe under identical measurement conditions [Fig. 2(d)]. Our calculations further show that the strongest Raman peak intensity of Te (the $A_1$ mode) is nearly two orders of magnitude larger than the $E_{2g}$ mode of MnTe under 633 nm excitation (Supplementary Note 4). This large intensity contrast highlights how Te-derived features can obscure the intrinsic Raman response of MnTe.

The same spectroscopic ambiguity is not limited to MnTe. Similar Raman peaks around 120 and 140 cm$^{-1}$ have been reported for many tellurides, including CdTe, CdMnTe, ZnTe, and GaTe [37-47]. Their nearly identical frequencies across host materials with different crystal structures and phonon spectra suggest that they are not independent intrinsic modes of each material. Instead, these peaks have been attributed to segregated Te clusters or precipitates [37-47].

Temperature-dependent Raman measurements were performed at a sample position where P1 was too weak to be resolved [Figs. 3(b)–3(d)]. Upon warming from 10 to 300 K, P2 progressively softens from approximately 98.5 to 92.8 cm$^{-1}$ and broadens, consistent with anharmonic phonon-phonon interactions, as described by the Balkanski anharmonic model [Fig. 3(d) and Supplementary Note 5] [48]. P3 remains near 101 cm$^{-1}$ over the temperature range where it can be resolved, while its intensity decreases with increasing temperature and becomes unresolvable above 85 K. P3 may be associated with local strain, which has been proposed as a possible origin of the Raman-mode splitting near 100 cm$^{-1}$ in MnTe [24]. Our findings suggest that only the P2 peak is intrinsic to MnTe, consistent with the $D_{6h}$ point group.

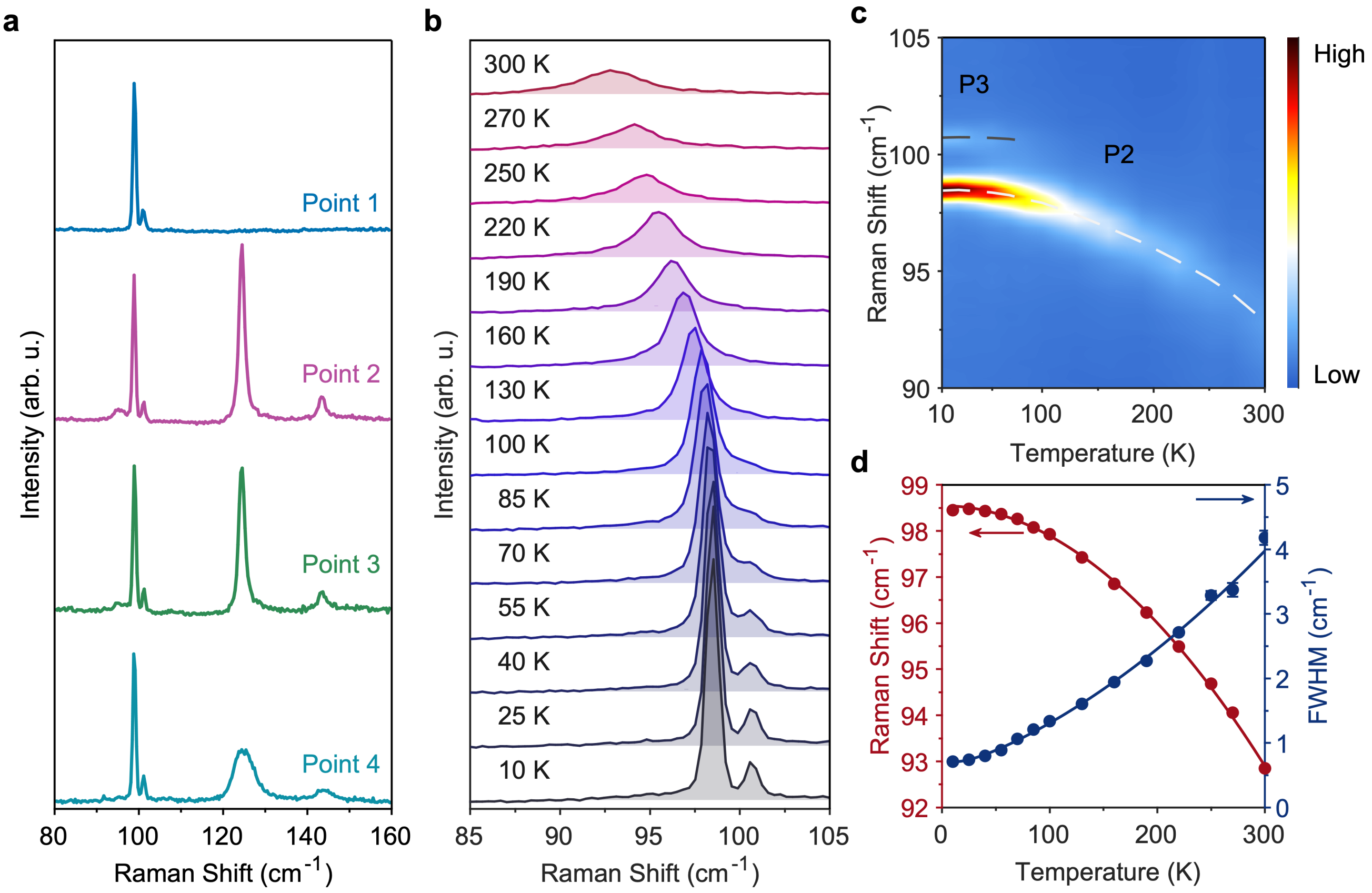

**Fig. 3 Spatial and temperature dependence of the Raman spectra in MnTe.** (a) Raman spectra collected at four positions at 1.8 K, normalized to P2 and vertically offset for clarity. (b) Raman spectra measured from 10 to 300 K, vertically offset for clarity. (c) Corresponding Raman intensity map; the white and gray dashed lines indicate the temperature evolution of P2 and P3, respectively. (d) Temperature dependence of the P2 peak frequency (red, left axis) and full width at half maximum (FWHM, blue, right axis). Solid curves are fits to the Balkanski anharmonic model [48].

Next, we further explore the crystal symmetry with SHG. SHG provides a sensitive probe of inversion symmetry, because the bulk electric-dipole (ED) susceptibility $\chi^{(2)}$ must vanish in a centrosymmetric crystal [49]. Accordingly, for the $D_{6h}$ point group, all bulk ED tensor elements vanish, whereas surface ED and bulk electric-quadrupole (EQ) contributions remain symmetry-allowed. For MnTe with the $P6_3/mmc$ space group, an ideal (0001) surface breaks the $6_3$ screw symmetry while preserving threefold rotation and three vertical mirror planes, resulting in $C_{3v}$ surface symmetry. The observed SHG can therefore arise from the coherent superposition of $C_{3v}$ surface ED and $D_{6h}$ bulk EQ fields. By contrast, the proposed noncentrosymmetric $D_{3h}$ structure allows bulk ED SHG [26]. The allowed tensor components and the resulting SHG signals are derived in Supplementary Note 6.

We measured SHG in reflection geometry on MnTe(0001) using 800-nm, 50-fs pulses incident at 45°, with the 400-nm signal analyzed in the S-out and P-out channels [Fig. 4(a)]. The incident-polarization angle $\alpha$ was defined from the P-polarization direction, while the sample azimuth $\varphi$ specifies the orientation of the crystallographic $a$ axis relative to the plane of incidence. At a fixed sample orientation ($\varphi = 0°$), we recorded the incident-polarization dependence at 4.8 and 323 K [Figs. 4(b,c)]. We also performed complementary rotational-anisotropy measurements at 300 K in the P-in/P-out configuration by rotating the sample about the $c$ axis [Fig. 4(d)].

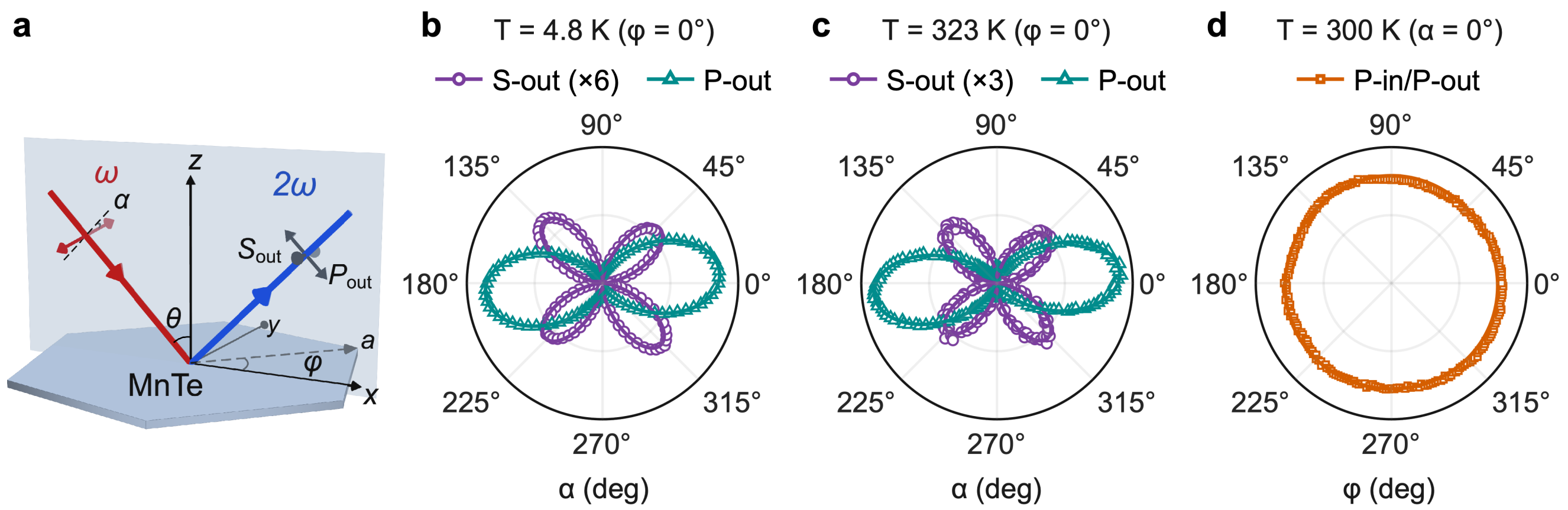

**Fig. 4 Polarization-resolved SHG response of MnTe in reflection geometry.** (a) Schematic of the oblique-incidence geometry with an incidence angle of $\theta = 45°$. The incident-polarization angle $\alpha$ is measured from the P-polarization direction, and $\varphi$ denotes the sample azimuth. The $x$ axis lies along the intersection of the plane of incidence and the sample surface, whereas $a$ denotes the crystallographic $a$ axis. The sample azimuth $\varphi$ is measured from $x$ to $a$. (b,c) Incident-polarization-dependent SHG intensities in the S-out and P-out channels at 4.8 and 323 K, respectively, with $\varphi = 0°$. (d) Sample-azimuth-dependent P-in/P-out SHG intensity at 300 K with $\alpha = 0°$. Symbols represent experimental data, and solid curves are fits to the $D_{6h}$ SHG model.

The incident-polarization scans at a fixed sample azimuth of $\varphi = 0°$ test the tensor selection rules in the S-out and P-out channels. In this geometry, the calculated response for a $D_{6h}$ bulk crystal, including the coherent superposition of $C_{3v}$ surface ED and bulk EQ fields, gives the following S-out and P-out intensities:

$$I_s(2\omega) \propto |C\sin 2\alpha - H_s(\tfrac{1}{2}\cos^2\alpha - \sin^2\alpha)|^2, \tag{1}$$

$$I_p(2\omega) \propto \left|A\cos^2\alpha + B\sin^2\alpha - \tfrac{1}{2}H_p\sin 2\alpha\right|^2, \tag{2}$$

where $A$, $B$, and $C$ are complex coefficients containing both surface ED and bulk EQ contributions, whereas $H_s$ and $H_p$ arise from a surface ED coefficient. The solid curves in Figs. 4(b) and 4(c) are fits to Eqs. (1, 2) for $D_{6h}$. At both 4.8 and 323 K, the measured patterns agree with the calculated responses and show no qualitative change across $T_N$. By contrast, the proposed $D_{3h}$ structure [26] gives the following pure bulk ED responses in the same geometry: $I_s(2\omega) \propto (\sin^2\alpha - \frac{1}{2}\cos^2\alpha)^2$ and $I_p(2\omega) \propto \sin^2 2\alpha$, which are completely different from the measured angular dependence in either channel [Fig. S7].

Rotational anisotropy in the P-in/P-out configuration provides an additional test to rule out the $D_{3h}$ bulk response. For the $D_{6h}$ symmetry, the intensity is $I_{pp}(2\omega,\varphi) \propto |A + \frac{H_p}{2\sqrt{2}}\sin 3\varphi|^2$. By contrast, the pure $D_{3h}$ bulk ED response requires a sixfold dependence, $I_{pp}(2\omega,\varphi) \propto \sin^2 3\varphi$. As shown in Fig. 4(d), the P-in/P-out intensity at 300 K remains nearly independent of $\varphi$, inconsistent with the sixfold pattern required by the pure $D_{3h}$ bulk ED model. A joint fit to the 300 K S-out and P-out polarization scans and the P-in/P-out rotational anisotropy scan shows a dominant isotropic amplitude and weak angular modulation [Supplementary Note 8 and Fig. S8]. These results are consistent with the combined $C_{3v}$ surface ED and $D_{6h}$ bulk EQ response, and support the preservation of bulk inversion symmetry in MnTe.

In addition, we calculated the incident-polarization and sample-azimuth dependences of the bulk ED SHG response allowed by the $D_3$ point group of Te [Supplementary Note 9]. The calculated results do not agree with the experimental data for MnTe, indicating that the SHG signals observed in MnTe are intrinsic.

In summary, we have distinguished the intrinsic Raman response of altermagnetic semiconductor MnTe from Te-derived spectral contributions. Several prominent Raman features, including a helicity-dependent mode and a feature previously attributed to two-magnon scattering, originate from segregated Te clusters or precipitates. Separating these contributions reveals a single intrinsic Raman-active mode in MnTe, consistent with the $D_{6h}$ point group symmetry. The SHG measurements indicate that the nonlinear signal arises from the surface ED and bulk EQ contributions, as expected for a centrosymmetric $D_{6h}$ crystal. These results revise the Raman assignments in MnTe and clarify how extrinsic Te contributions affect the interpretation of lattice excitations. The evidence for preserved bulk inversion symmetry provides an essential basis for further investigations of its symmetry-dependent electronic and magnetic properties.

# Methods

## Sample preparation

MnTe single crystals were prepared using Sb as the flux, following the previously reported procedure [50]. Mn plates (99.9%, Macklin), Te powders (99.99%, Macklin), and Sb blocks (99.999%, Macklin) were combined in a Mn:Te:Sb molar ratio of 1:1:20. The mixture was placed in an alumina crucible and enclosed in an evacuated quartz ampoule. The furnace temperature was raised to 1050 °C over 15 h and maintained at this temperature for 10 h. The melt was then cooled to 700 °C at 2 °C $\mathrm{h}^{-1}$, at which point the excess Sb flux was separated from the crystals by high-speed centrifugation. This procedure yielded black, shiny crystals with typical dimensions of $5 \times 4 \times 1\,\mathrm{mm}^3$.

Te single crystals were grown in an evacuated quartz ampoule using a two-zone tube furnace. High-purity Te pieces (99.999%) were annealed at 400 °C for 24 h in a sealed high-purity $\mathrm{H}_2$ atmosphere to remove surface oxides. One gram of the treated Te pieces was sealed in an evacuated quartz ampoule ($\sim 10^{-4}$ Pa), which was placed horizontally in the furnace. Both zones were heated to 450 °C, after which the source zone was maintained at 450 °C while the growth zone was gradually cooled to 360 °C. The relatively small temperature gradient helped control the nucleation rate and was crucial for the growth of centimeter-scale Te single crystals. The source and growth zones were then maintained at 450 and 360 °C, respectively, for 5 days to allow further crystal growth, after which the furnace was allowed to cool naturally to room temperature. A centimeter-scale Te single crystal with naturally developed $(h00)$ facets was obtained in the growth zone, with dimensions of $10 \times 3.2 \times 1.5\,\mathrm{mm}^3$.

To obtain the $(00l)$ surfaces, the as-grown Te crystal was cut perpendicular to the $(h00)$ facets using a precision diamond-wire saw. The resulting single-crystal plate had dimensions of $3.2 \times 1.5 \times 0.8\,\mathrm{mm}^3$ and was subsequently mechanically polished to obtain flat, mirror-like $(00l)$ surfaces. Basic characterization of the MnTe and Te single crystals is presented in Supplementary Note 1.

## Raman spectroscopy

As shown schematically in Fig. S1(a), Raman measurements were performed with the sample mounted in an optical closed-cycle helium cryostat over the temperature range of 1.8–300 K. Spectra were acquired in a backscattering configuration using a micro-Raman spectrometer equipped with a charge-coupled device (CCD) detector. A 632.8 nm He–Ne laser was used as the excitation source. The laser plasma lines and the Rayleigh line were suppressed using appropriate filters. The laser beam was focused onto the sample at normal incidence via a 40× microscope objective lens to a spot size of about 2 μm. The backscattered signal was dispersed using a grating with 1800 grooves $\mathrm{mm}^{-1}$, yielding a spectral resolution of about 0.35 $\mathrm{cm}^{-1}$. To minimize laser-induced heating while maintaining an adequate signal-to-noise ratio, the laser power on the sample was kept below 1 mW.

For circular-polarization-resolved measurements, the linearly polarized incident beam was converted into right- or left-circularly polarized light (R or L) using a quarter-wave plate. The backscattered light passed through the same quarter-wave plate, followed by a linear polarizer and a half-wave plate before entering the spectrometer. The linear polarizer was used to select the desired polarization component, while the half-wave plate maintained a fixed polarization orientation of the light entering the spectrometer. This arrangement allowed us to select the desired polarization configurations. In the polarization notation used here, the first and second letters denote the polarizations of the incident and analyzed scattered light, respectively. LL and RR denote the co-circular configurations, whereas RL and LR denote the cross-circular configurations. Unless otherwise specified, Raman spectra were acquired with linearly polarized excitation and without polarization analysis of the scattered light.

## Second-harmonic generation

As shown schematically in Fig. S1(b), second-harmonic generation (SHG) measurements were performed in reflection geometry using linearly polarized femtosecond pulses from a Ti:sapphire amplifier (800 nm central wavelength, 50 fs pulse duration, 10 kHz repetition rate). Before reaching the sample, the incident beam passed sequentially through a Glan–Taylor polarizer to define a linear polarization, a motorized half-wave plate to adjust its polarization angle, and an 800 nm band-pass filter. The beam was then focused at an incidence angle of 45° using a 100 mm focal-length lens, producing a spot diameter of ~34 μm. The average incident power at the sample was 400 μW. The reflected light was collimated and passed through one 400 nm band-pass filter and two 450 nm short-pass filters to suppress the residual fundamental beam before detection. The filtered 400 nm SHG signal was detected with a photomultiplier tube (PMT), whose output current was converted to a voltage by a current preamplifier and recorded with a lock-in amplifier referenced to the laser repetition frequency.

Two complementary SHG protocols were used. In polarization-dependent measurements, the sample was mounted in an optical closed-cycle helium cryostat with an operating temperature range of 4–400 K. The plane of incidence was perpendicular to the crystallographic $a$–$b$ plane. The polarization-dependent measurements were performed at a fixed sample azimuth of $\varphi = 0°$, with the intersection of the plane of incidence and the sample surface parallel to the crystallographic $a$ axis. The incident polarization angle was varied with the motorized half-wave plate. A second half-wave plate rotated the SHG polarization before a fixed linear analyzer to select $p$- or $s$-polarized output. In rotational-anisotropy measurements, the sample was mounted on a motorized rotation stage and measured only at room temperature. The incident and analyzed polarizations were fixed in the $p$-in/$p$-out configuration, and the sample was rotated about its surface normal to vary the sample azimuth, defined as the angle between the crystallographic $a$ axis and the intersection of the plane of incidence with the sample surface. Throughout this work, $p$ ($s$) denotes the electric-field component parallel (perpendicular) to the plane of incidence.

## First-principles calculations

The phonon spectra of MnTe were computed within density-functional perturbation theory (DFPT) [51] as implemented in the Vienna *ab initio* simulation package (VASP) [52, 53], combined with the PHONOPY package [54] for post-processing. Exchange and correlation were described by the generalized gradient approximation of Perdew, Burke, and Ernzerhof (PBE) [55], and the projector-augmented-wave (PAW) method [53, 56] was employed. The antiferromagnetic (AFM) order was treated within spin-polarized calculations without SOC and $U$. A $3 \times 3 \times 2$ supercell and a $3 \times 3 \times 3$ $k$-point mesh were used, and the plane-wave energy cutoff was set to 500 eV.

The Raman tensor of a phonon mode is obtained from the derivative of the macroscopic dielectric function with respect to the normal coordinate of that mode. For a displacement $\pm \boldsymbol{u}_b$ of atom $i$ along Cartesian direction $\alpha$ (here $b$ is the composite index of $i$ and $\alpha$), the finite-difference estimate

$$R_{ij}(b) = \frac{\partial \varepsilon_{ij}}{\partial u_b} = \frac{\varepsilon_{ij}^{+u_b}(\omega_L) - \varepsilon_{ij}^{-u_b}(\omega_L)}{2|u_b|}, \tag{3}$$

is evaluated from two independent calculations with positive and negative displacements of magnitude $|u_b| = 0.005$ Å.

For a phonon mode with (mass-weighted) eigenvector coefficients $c_b = \mathrm{Re}\, c_b + i\, \mathrm{Im}\, c_b$, normalized as $\sum_b |c_b|^2 = 1$, the mode Raman tensor follows from the linear combination

$$R_{ij}^{\mathrm{mode}} = \sum_b \frac{c_b R_{ij}(b)}{\sqrt{m_b}}, \tag{4}$$

where $m_b$ is the mass of the displaced atom. The intensity of Raman scattering in the polarization channel $\alpha \rightarrow \beta$ (incident polarization $\boldsymbol{e}_\alpha$, scattered polarization $\boldsymbol{e}_\beta$) is

$$I_{\alpha\beta} = |\boldsymbol{e}_\beta^* \cdot R^{\mathrm{mode}} \cdot \boldsymbol{e}_\alpha|^2. \tag{5}$$

Linear channels follow from Cartesian unit vectors, while the circular channels are defined with the helicity basis $\boldsymbol{e}_{\mathrm{R,L}} = (\hat{\boldsymbol{x}} \pm i\hat{\boldsymbol{y}})/\sqrt{2}$ (referred to the $+z$ propagation direction), yielding $I_{\mathrm{RR}}$, $I_{\mathrm{LL}}$, $I_{\mathrm{RL}}$, and $I_{\mathrm{LR}}$.

The dielectric tensors entering Eq. (3) were computed with the independent-particle approach as implemented in VASP [52, 57]. Both MnTe and Te were treated at the PBE/PAW level with a plane-wave cutoff of 500 eV, a Γ-centered 24 × 24 × 16 *k*-point mesh for the primitive cells, and SOC was not considered. For MnTe, all the dielectric functions were calculated for the collinear AFM ground state without *U*.

Raman intensities were evaluated at a photon energy of 1.959 eV (633 nm) by extracting $\varepsilon_{ij}(\omega_L)$ from each displaced-structure calculation. The calculated spectra were constructed using Lorentzian functions with a common full width at half maximum of 4 $\mathrm{cm}^{-1}$ for MnTe and Te. This broadening was introduced solely for visualization and does not represent calculated physical phonon linewidths.

## Acknowledgements

The work was supported by the National Key R&D Program of China (Grants No. 2025YFA1411200), the Beijing Natural Science Foundation (Grants No. Z240006), and the National Natural Science Foundation of China (Grants No. 12361141826, No. 12421004, No. 12127807 and No. 12322401). This work was supported by the Synergetic Extreme Condition User Facility (SECUF, https://cstr.cn/31123.02.SECUF).

## References

(1) Ma, H.-Y.; et al. Multifunctional antiferromagnetic materials with giant piezomagnetism and noncollinear spin current. *Nat. Commun.* **2021**, *12* (1), No. 2846.

(2) Liu, P.; et al. Spin-group symmetry in magnetic materials with negligible spin-orbit coupling. *Phys. Rev. X* **2022**, *12* (2), No. 021016.

(3) Šmejkal, L.; Sinova, J.; Jungwirth, T. Beyond conventional ferromagnetism and antiferromagnetism: A phase with nonrelativistic spin and crystal rotation symmetry. *Phys. Rev. X* **2022**, *12* (3), No. 031042.

(4) Šmejkal, L.; Sinova, J.; Jungwirth, T. Emerging research landscape of altermagnetism. *Phys. Rev. X* **2022**, *12* (4), No. 040501.

(5) Šmejkal, L.; et al. Anomalous Hall antiferromagnets. *Nat. Rev. Mater.* **2022**, *7* (6), 482–496.

(6) Bai, L.; et al. Altermagnetism: Exploring new frontiers in magnetism and spintronics. *Adv. Funct. Mater.* **2024**, *34* (49), No. 2409327.

(7) Liu, Y.; et al. Symmetry classification of magnetic orders using oriented spin space groups. *Nature* **2026**, *652* (8111), 869–873.

(8) Chen, X.; et al. Unconventional magnons in collinear magnets dictated by spin space groups. *Nature* **2025**, *640* (8058), 349–354.

(9) Jungwirth, T.; et al. Symmetry, microscopy and spectroscopy signatures of altermagnetism. *Nature* **2026**, *649* (8098), 837–847.

(10) Jungwirth, T.; et al. Altermagnetic spintronics. *Nat. Phys.* **2026**, *22* (7), 1012–1021.

(11) Song, C.; et al. Altermagnets as a new class of functional materials. *Nat. Rev. Mater.* **2025**, *10* (6), 473–485.

(12) Gonzalez Betancourt, R. D.; et al. Spontaneous anomalous Hall effect arising from an unconventional compensated magnetic phase in a semiconductor. *Phys. Rev. Lett.* **2023**, *130* (3), No. 036702.

(13) Liebman-Peláez, A.; et al. Strain continuously rotates the Néel vector in altermagnetic MnTe. arXiv:2604.07653, **2026**.

(14) Hubert, M.; et al. Anomalous spectroscopical effects in an antiferromagnetic semiconductor: The case of magneto-optical Kerr effect. *Phys. Status Solidi B* **2025**, *262* (7), No. 2400541.

(15) Krempaský, J.; et al. Altermagnetic lifting of Kramers spin degeneracy. *Nature* **2024**, *626* (7999), 517–522.

(16) Amin, O. J.; et al. Nanoscale imaging and control of altermagnetism in MnTe. *Nature* **2024**, *636* (8042), 348–353.

(17) Liu, Z.; et al. Chiral split magnon in altermagnetic MnTe. *Phys. Rev. Lett.* **2024**, *133* (15), No. 156702.

(18) Gray, I.; et al. Time-resolved magneto-optical effects in the altermagnet candidate MnTe. *Appl. Phys. Lett.* **2024**, *125* (21), No. 212404.

(19) Zhu, C.; et al. Ultrafast dynamics of optically excited charge carriers in the room-temperature antiferromagnetic semiconductor α-MnTe. *Phys. Rev. Mater.* **2023**, *7* (5), No. 054601.

(20) Bossini, D.; et al. Femtosecond phononic coupling to both spins and charges in a room-temperature antiferromagnetic semiconductor. *Phys. Rev. B* **2021**, *104* (22), No. 224424.

(21) Yang, X.; et al. Ultrafast magneto-optical fingerprints of altermagnetism in MnTe. *Phys. Rev. Lett.* **2026**, *136* (25), No. 256702.

(22) Mobasser, S. R.; Hart, T. R. Raman scattering from phonons and magnons in magnetic semiconductor, MnTe. In *Spectroscopic Characterization Techniques for Semiconductor Technology II*; SPIE, **1985**; Vol. 524, pp 137–144.

(23) Zhang, J.; et al. Spin-phonon coupling and two-magnons scattering behaviors in hexagonal NiAs-type antiferromagnetic MnTe epitaxial films. *J. Raman Spectrosc.* **2020**, *51* (8), 1383–1389.
(24) Uykur, E.; et al. Revisiting the symmetry and optical phonons of altermagnetic α-MnTe. *Phys. Rev. B* **2026**, *113* (22), No. 224107.
(25) Sheokand, H.; et al. Observation of the optical phonons in α-MnTe films. arXiv:2605.26075, **2026**.
(26) Wu, A.; et al. Domains with noncentrosymmetric structural distortion in altermagnetic MnTe. *Phys. Rev. B* **2026**, *114* (7), No. 074436.
(27) Thapa, B.; Belashchenko, K. D.; Mazin, I. I. Possible origin of the 175 $cm^{-1}$ Raman mode in altermagnetic MnTe. *Phys. Rev. B* **2026**, *114* (7), No. 074416.
(28) Azam, N.; et al. The origin of the observed Raman peaks in α-MnTe. arXiv:2609.19467, **2026**.
(29) Bhardwaj, P.; et al. Nature of the 175–180 $cm^{-1}$ Raman feature in altermagnetic α-MnTe. arXiv:2609.39108, **2026**.
(30) Ishito, K.; et al. Chiral phonons: Circularly polarized Raman spectroscopy and ab initio calculations in a chiral crystal tellurium. *Chirality* **2023**, *35* (6), 338–345.
(31) Zhang, T.; et al. Weyl phonons in chiral crystals. *Nano Lett.* **2023**, *23* (16), 7561–7567.
(32) Pine, A. S.; Dresselhaus, G. Raman spectra and lattice dynamics of tellurium. *Phys. Rev. B* **1971**, *4* (2), 356–371.
(33) Mazin, I. I. Altermagnetism in MnTe: Origin, predicted manifestations, and routes to detwinning. *Phys. Rev. B* **2023**, *107* (10), No. L100418.
(34) Mazin, I. I.; Belashchenko, K. D. Origin of the gossamer ferromagnetism in MnTe. *Phys. Rev. B* **2024**, *110* (21), No. 214436.
(35) Takahashi, K.; et al. Symmetry and minimal Hamiltonian of nonsymmorphic collinear antiferromagnet MnTe. *npj Quantum Mater.* **2025**, *10*, No. 70.
(36) Zhang, S.; et al. Comprehensive study of phonon chirality under symmetry constraints. *Phys. Rev. B* **2026**, *113* (2), No. 024302.
(37) Manjón, F. J.; et al. Anomalous Raman modes in tellurides. *J. Mater. Chem. C* **2021**, *9* (19), 6277–6289.
(38) Zitter, R. N. Raman detection of tellurium layers on surfaces of CdTe. *Surf. Sci.* **1971**, *28* (1), 335–338.
(39) Shin, S. H.; et al. Characterization of Te precipitates in CdTe crystals. *Appl. Phys. Lett.* **1983**, *43* (1), 68–70.
(40) Amirtharaj, P. M.; Pollak, F. H. Raman scattering study of the properties and removal of excess Te on CdTe surfaces. *Appl. Phys. Lett.* **1984**, *45* (7), 789–791.
(41) Sochinskii, N. V.; et al. Effect of thermal annealing on Te precipitates in CdTe wafers studied by Raman scattering and cathodoluminescence. *J. Appl. Phys.* **1995**, *77* (6), 2806–2808.
(42) Luan, L.-j.; et al. Studies on Raman scattering and Te precipitates in cadmium manganese telluride. *J. Alloys Compd.* **2009**, *477* (1-2), 399–402.
(43) Azhniuk, Y. M.; et al. Precipitates of selenium and tellurium in II-VI nanocrystal-doped glass probed by Raman scattering. *Phys. Status Solidi B* **2011**, *248* (3), 674–679.
(44) Lindberg, G. P.; et al. Effects of pressure on photo-induced formation of Se and Te clusters in II-VI compounds. *Phys. Status Solidi B* **2013**, *250* (4), 711–715.
(45) Zhang, L.; et al. Growth and characterization of highly nitrogen doped ZnTe films on GaAs (001) by molecular beam epitaxy. *Mater. Sci. Semicond. Process.* **2015**, *29*, 351–356.
(46) Cuenca-Gotor, V. P.; et al. Structural, vibrational, and electronic study of α-$As_2Te_3$ under compression. *J. Phys. Chem. C* **2016**, *120* (34), 19340–19352.

(47) Ananth Kumar, R. T.; et al. Scrutiny of structural disorder using Raman spectra and Tauc parameter in $GeTe_2$ thin films. *J. Phys.: Conf. Ser.* **2017**, *869* (1), No. 012018.
(48) Balkanski, M.; Wallis, R. F.; Haro, E. Anharmonic effects in light scattering due to optical phonons in silicon. *Phys. Rev. B* **1983**, *28* (4), 1928–1934.
(49) Fiebig, M.; Pavlov, V. V.; Pisarev, R. V. Second-harmonic generation as a tool for studying electronic and magnetic structures of crystals: Review. *J. Opt. Soc. Am. B* **2005**, *22* (1), 96–118.
(50) Gao, B.; et al. Electrodynamics of a prototypical altermagnetic compound MnTe. *Chin. Phys. B* **2026**, *35* (3), No. 037801.
(51) Baroni, S.; et al. Phonons and related crystal properties from density-functional perturbation theory. *Rev. Mod. Phys.* **2001**, *73*, 515–562.
(52) Kresse, G.; Furthmüller, J. Efficient iterative schemes for ab initio total-energy calculations using a plane-wave basis set. *Phys. Rev. B* **1996**, *54*, 11169–11186.
(53) Kresse, G.; Joubert, D. From ultrasoft pseudopotentials to the projector augmented-wave method. *Phys. Rev. B* **1999**, *59*, 1758–1775.
(54) Togo, A.; Tanaka, I. First principles phonon calculations in materials science. *Scr. Mater.* **2015**, *108*, 1–5.
(55) Perdew, J. P.; Burke, K.; Ernzerhof, M. Generalized gradient approximation made simple. *Phys. Rev. Lett.* **1996**, *77*, 3865–3868.
(56) Blöchl, P. E. Projector augmented-wave method. *Phys. Rev. B* **1994**, *50*, 17953–17979.
(57) Gajdoš, M.; et al. Linear optical properties in the projector-augmented-wave methodology. *Phys. Rev. B* **2006**, *73*, No. 045112.